\documentclass[aps,twocolumn]{revtex4}
\usepackage{graphicx}
\usepackage{tabularx}

\begin{document}

\title{Comment on N. Rijal {\em et al.} ``Measurement of d + $^7$Be Cross Sections \\ for Big-Bang Nucleosynthesis".}

\author{Moshe Gai}
\affiliation{LNS at Avery Point, University of Connectciut, 1084 Shennecossett Rd., Groton, CT 06340}

\begin{abstract}
\      \\
Rijal {\em et al.} in their recent publication [Phys. Rev. Lett {\bf 122}, 182701 (2019)], on ``Measurement of d + $^7$Be Cross Sections for Big-Bang Nucleosynthesis (BBN)", misrepresent their result, they misrepresent previous work of Parker (72) and of Caughlan and Fowler (88), and quite possibly, contradicts the very BBN theory that has been established over the last few decades. This comment is intended to correct these misrepresentations and critically review their claims on BBN.
\end{abstract}

\preprint{UConn-40870-00XX}

\maketitle

\begin{figure}
 \includegraphics[width=3in]{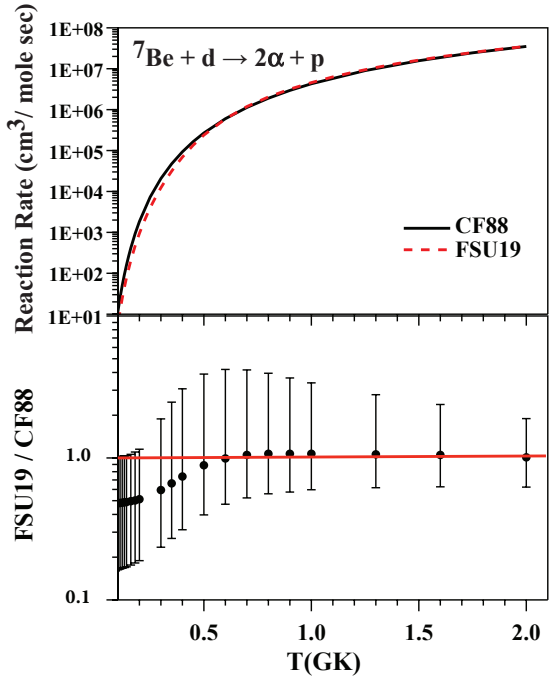}
 \caption{\label{Rate} (Color Online) A comparison of the FSU19 reaction rate published in \cite{PRL} and the CF88 rate \cite{CF88}, and the ratio of the two rates. Over the BBN region of interest of 0.5 - 0.9 GK, the two rates are identical.}
\end{figure}

\begin{figure}
 \includegraphics[width=3in]{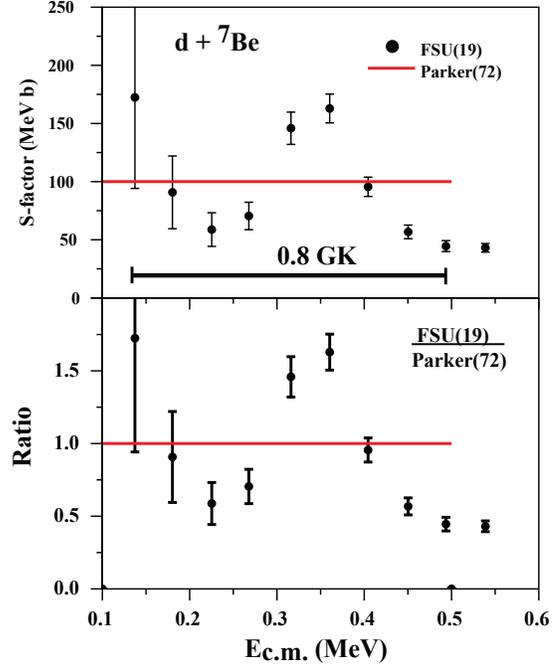}
 \caption{\label{Rate} (Color Online) A comparison of the s-factor measured in the Gamow window of BBN indicated in \cite{PRL} and Parker's 1972 (educated guess) s-factor of $\sim$100 MeV b \cite{Parker}, and the ratio of the two s-factors.}
\vspace{-0.5cm}
\end{figure}

Rijal {\em et al.} \cite{PRL} measured the cross section of the d + $^7$Be $\rightarrow \  2\alpha$ + p reaction at energies relevant for Big Bang Nucleosynthesis (BBN) from which they deduced the new ``rates derived from our excitation function" \cite{PRL},  shown in Fig. 5 of \cite{PRL}, hereafter the new ``FSU19 rate". Based on their new rate they claim that ``the resonance
reduces the predicted abundance of primordial $^7$Li", and they derive: ``our reaction rates predict ($^7$Li/H)$_p$ = 4.24 - 4.61$\times$ 10$^{-10}$", as shown in Fig. 6 of \cite{PRL}. 

In this comment we demonstrate that the FSU19 rate of the d + $^7$Be reaction, is the same rate that has been used in BBN for over fourty years. And the $^7$Li abundance deduced in \cite{PRL} was published more than twenty years ago. Furthermore, the impression that the FSU19 rate is larger (which leads to smaller primordial $^7$Li abundance) is based on a selective ``straw man" comparison with other rates that have not been used in BBN. Specifically, their statement that "the resonance reduces the predicted abundance of primordial $^7$Li", is incorrect. No reduction of the abundance of $^7$Li beyond that which was already calculated by the practitioner of BBN, can be deduced from the FSU19 rate.

Rijal {\em et al.} chose to compare in Fig. 5 \cite{PRL}, their d + $^7$Be rate to a rate based on s-factor data that was obtained at higher energies by Kavanagh in 1960 \cite{Kavanagh} and the more recent rate published by Angulo {\em et al.} \cite{LLN}. These comparisons give the impression that a new higher d +$^7$Be rate was measured in the FSU experiment. As such they conclude that their new rate including a resonance ``reduces the predicted abundance of primordial $^7$Li". But the so-labeled ``Kavanagh rate", was never used by the practitioner of BBN (over the last sixty years) and it is not relevant for the discussion of BBN. The rate of Angulo {\em et al.}, was also not used in BBN, since it is even smaller than the d + $^7$Be rate that was used in BBN. Simply put, the ``straw man" comparison of the FSU19 rate with the rates of Kavanagh and Angulo {\em et al.} shown in Fig. 5 of \cite{PRL}, is irrelevant for BBN.
  
Instead, in Fig. 1 we show the d + $^7$Be rate that was used by the practitioner in the field of BBN over the last thirty years as already published in 1988 by Caughlan and Fowler, hereafter the ``CF88 rate" \cite{CF88}. Prior to CF88 the CFZ75 compilation of 1975 \cite{CFZ} listed the same rate as in CF88. Rijal {\em et al.} did not consider CF88 (or CFZ75 rate) and in their Fig. 5 they chose not to compare the CF88 rate to their rate . In Fig. 1 we compare the FSU19 rate to the CF88 rate. The ratio of the two rates is also shown in Fig. 1. Clearly, over the region of interest for BBN indicated in Fig. 5 of \cite{PRL}, of T= 0.8 GK, in of itself incorrect, it should read 0.5 - 0.9 GK, the central values of the FSU19 rate is identical to the CF88 rate. Rijal {\em et al.} did not measure a new rate for the d + $^7$Be interaction during BBN. 

Furthermore, considering the very large uncertainties of the FSU19 rate \cite{PRL} shown in Fig. 1, which is due to the ill determined resonance energy, rates which differ by up to a factor of 10 are consistent with their data. As such the CF88 and FSU19 rates are hardly different if not identical over the entire reported temperature range of Fig. 5. 

Since no state is known in $^9$B at the proposed "new resonance" energy of 16.85 MeV, resolving such a major systematical uncertainty would have been essential for making the FSU19 rate of some use for BBN, beyond merely confirming the CF88 rate. Instead of resolving this systematic uncertainty Rijal {\em et al.} resort to speculations that their resonance may have been observed in another measurement of the ($^3$He,t) reaction at 16.80 MeV.

The seminal compilation of stellar nuclear rates by Caughlan and Fowler includes the interaction of d + $^7$Be that is used today in BBN. Using the CF88 rate it was concluded long ago that the d + $^7$Be reaction does not play a significant role in BBN. As such it is not included in the twelve canonical reactions that are relevant for BBN \cite{BBN1}.  We note that the more recent REACLIB compilation \cite{reaclib}, referenced by Rijal {\em et al.}, contains no new information on the interaction of d + $^7$Be beyond the CF88 compilation. In fact, quite to the contrary, in REACLIB we find the list of the well known twelve canonical reactions of BBN (listed as ``popular reactions of BBN") that does not include the d +$^7$Be reaction. This reaction was recognized as irrelevant for BBN long before the compilation of the REACLIB.

The d + $^7$Be rate listed in the CF88 compilation, was based on the work of Parker \cite{Parker}, which Rijal {\em et al.} dismiss as arbitrary for multiplying the ``cross-section data from Kavanagh \cite{Kavanagh} by an arbitrary factor of 3". But Parker stated  very clearly that he ``multiplied by a factor of $\sim$3 [Kavanagh's s-factor] to take into account contributions from higher excited states in $^8$Be". Parker's educated guess of the s-factor was not ``arbitrary" and should not be labeled as such, even if one claims that it is based on incorrect assumption(s).

In Fig. 2, we compare the s-factor of the d + $^7$Be reaction deduced by Parker ($\sim 100$ MeV b) \cite{Parker} with the FSU19 s-factor \cite{PRL} measured at their indicated region of interest (0.8 GK). In this region of interest, Parker's simple assumption of a constant s-factor is on average a reasonable approximation of the s-factor measured by the FSU group. It is than little wonder that Rijal {\em et al.} also observed that the ($^7$Be/H)$_p \ \approx 4.51 \times \ 10^{-10}$, predicted using Parker's rate, agrees with their result, and indeed with their entire stellar burning rate.

We conclude that Parker's rate for the d + $^7$Be reaction which is based on the most elementary assumption of a constant s-factor, the CF88 (and CFZ75) rates which are based on Parker's rate, and the FSU19 rate, are all the same over the region of interest, and no new information relevant for BBN has been gained since Parker's paper of 1972. The conclusion that the interaction of d + $^7$Be plays no significant role in BBN, was not altered by Rijal {\em et al.} As such, the d + $^7$Be interaction must still be ignored and not included in the tweleve canonical reactions of BBN \cite{BBN1}. 

Perhaps a a more disturbing observation is the discrepancy between the FSU findings and our well established understanding of BBN. For example, using the publicly available PRIMAT code \cite{Coc}, the d + $^7$Be rate would have to be $\sim$30 times larger than the CF88 rate to play a significant role in BBN. Similarly, not including the Parker rate (or the FSU19 rate) the practitioner of BBN \cite{Ken} concluded long ago a change in the $^7$Li/H abundance by no more than 1\%. It is not clear why Rijal {\em et al.} claim a Li/H change by 15\%. We can only suspect a mistake in their unspecified code for BBN calculation. Simply put, Rijal {\em et al.} \cite{PRL}, contradict the very understanding of BBN that has been established over many years of research by the practitioners in the field. Indeed, the ($^7$Li/H)$_p$ quoted by Rijal {\em al.} of 4.24 -  4.61 $\times ^{-10}$ can already be deduced, when using the now known baryon density, from the BBN calculations of more than twenty years ago \cite{BBN1,BBN2}.

In conclusion, the long standing observation that the interaction of d + $^7$Be plays very little role if any, in BBN, has not been altered using the so-called FSU19 new rate  \cite{PRL}. And in any case, the FSU19 rate published in \cite{PRL}, should not be considered anew, since the same rate was already published thirty years ago by Caughlan and Fowler \cite{CF88}.
\vspace{-0.5cm}

\section{Acknowledgement}
The author wish to acknowledge discussions with I. Wiedenhoever and J. C. Blackmon, the co-authors of \cite{PRL}. Helpful comments from Alain Coc and Kenneth Nollett on their BBN calculations are acknowledged. The material presented in this paper is based upon work supported by the U.S. Department of Energy Award
Number DE-FG02-94ER40870.

\end{document}